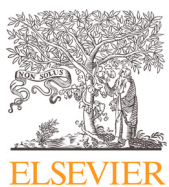

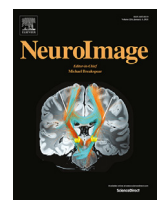

# Cascaded Multi-Modal Mixing Transformers for Alzheimer's Disease Classification with Incomplete Data ☆

Linfeng Liu [a,*], Siyu Liu [b], Lu Zhang [a,b], Xuan Vinh To [a,1], Fatima Nasrallah [a], Shekhar S. Chandra [b]

[a] Queensland Brain Institute, The University of Queensland, Australia
[b] School of Information Technology and Electrical Engineering, The University of Queensland, Australia

## A R T I C L E   I N F O

## A B S T R A C T

Accurate medical classification requires a large number of multi-modal data, and in many cases, different feature types. Previous studies have shown promising results when using multi-modal data, outperforming single-modality models when classifying diseases such as Alzheimer's Disease (AD). However, those models are usually not flexible enough to handle missing modalities. Currently, the most common workaround is discarding samples with missing modalities which leads to considerable data under-utilisation. Adding to the fact that labelled medical images are already scarce, the performance of data-driven methods like deep learning can be severely hampered. Therefore, a multi-modal method that can handle missing data in various clinical settings is highly desirable. In this paper, we present Multi-Modal Mixing Transformer (3MT), a disease classification transformer that not only leverages multi-modal data but also handles missing data scenarios. In this work, we test 3MT for AD and Cognitively normal (CN) classification and mild cognitive impairment (MCI) conversion prediction to progressive MCI (pMCI) or stable MCI (sMCI) using clinical and neuroimaging data. The model is based on a novel Cascaded Modality Transformers architecture with cross-attention to incorporate multi-modal information for more informed predictions. We propose a novel modality dropout mechanism to ensure an unprecedented level of modality independence and robustness to handle missing data scenarios. The result is a versatile network that enables the handling of arbitrary numbers of modalities with different feature types and also ensures full data utilization in missing data scenarios. The model is trained and evaluated on the Alzheimer's Disease Neuroimaging Initiative (ADNI) dataset with the state-of-the-art performance and further evaluated with The Australian Imaging Biomarker & Lifestyle Flagship Study of Ageing (AIBL) dataset with missing data.

## 1. Introduction

Alzheimer's Disease (AD) is a chronic neurodegenerative disease characterized by the gradual death of nerve cells and the loss of brain tissue. Over time, AD progresses, leading to a decline in the patient's ability to function independently. The initial onset of AD is typically slow, with symptoms worsening as the disease advances (McKhann et al., 1984). Projections indicate that the number of affected individuals will double within the next two decades, meaning that by 2025, approximately one in 85 people will suffer from AD (Alzheimer's Association, 2019). Unfortunately, there are currently no known medical procedures to reverse the progression of AD, and clinical trials investigating potential treatments have shown limited improvements (Servick, 2019). Conse-quently, early detection of AD signs is crucial for the well-being of patients.

AD diagnosis often relies on various types of data collected from different procedures, including cognitive examinations such as Mini-Mental State Examination (MMSE) or neuropsychological testing (McKhann et al., 2011), Positron emission tomography (PET) such as Pittsburgh compound B-PET ($A\beta$), Flortaucipir-PET (tau) and Fluorodeoxyglucose (FDG)-PET for detecting clear AD hallmarks such as amyloid and tau, respectively (Hanseeuw et al., 2019; Small et al., 2006), and biomarkers in the Cerebrospinal Fluid (CSF) to inform biological processes (Humpel, 2011; Meredith et al., 2013). In terms of accuracy, according to Beach et al. (2012), an average clinician's AD

---

☆ Data used in the preparation of this article was obtained from the Alzheimer's Disease Neuroimaging Initiative (ADNI) database (adni.loni.usc.edu) and the Australian Imaging Biomarkers and Lifestyle flagship study of ageing (AIBL) funded by the Commonwealth Scientific and Industrial Research Organisation (CSIRO) which was made available at the ADNI database (www.loni.usc.edu/ADNI). The ADNI and AIBL researchers contributed data but did not participate in analysis or writing of this report.
* Corresponding author.
 *E-mail address:* linfeng.liu@uq.edu.au (L. Liu).
[1] Data preprocessing and Neuroimaging guidance

https://doi.org/10.1016/j.neuroimage.2023.120267.
Received 20 February 2023; Received in revised form 27 June 2023; Accepted 5 July 2023
Available online 7 July 2023.

classification sensitivity is between 70.9% to 87.3% while the specificity is only between 44.3% and 70.8% indicating significant room for improvement. To aid clinicians in more accurately diagnosing AD, computer-aided methods like machine learning have been increasingly adopted for analyzing multi-modal medical data associated with AD (Klöppel et al., 2008; Zhang et al., 2015). In classic machine learning, imaging features extracted from medical imaging data such as Magnetic Resonance Imagings (MRIs), are commonly applied to analyze the brain volume change from Cognitively normal (CN) to AD. Whole brain voxel-wise volume, hippocampus volume, and cortical thickness measurements are usually extracted during the image pre-processing stage (Hutton et al., 2009). Then these extracted features are fed into classifiers such as random forests (Lebedev et al., 2014) and support vector machines (Klöppel et al., 2008). Recently, deep learning approaches have been gradually adopted for AD studies. The main advantage of deep learning is its data-driven nature. Typically, deep learning models automatically learn from the data in an end-to-end manner, and the pipeline involves little to no human intervention or feature engineering. As a result, deep learning methods offer a high degree of flexibility in their functionality, allowing for the application of popular design archetypes tailored to different data types. For instance, Convolutional Neural Networks (CNNs) are commonly employed for analyzing imaging data, while transformers have emerged as effective models for handling time series data. These design choices have gained popularity due to their ability to effectively capture and exploit the specific characteristics and patterns present in each data type.

When applying deep learning to AD analyze, it is often beneficial to incorporate multi-modal data as single modalities cannot usually yield the best performance. Guan et al. (2021); Qiu et al. (2020) have attempted combining imaging data and clinical data where the imaging part uses a CNN and the clinical information are injected into the model to provide more context. However, these models lack the flexibility to handle missing data scenarios which are common in multi-modal studies. For example, a model trained with imaging data, MMSE and age are not capable of predictions with MMSE only. Hence, the most common workaround has been discarding samples with missing modalities at the risk of data underutilization. This problem calls for a more versatile approach that can automatically handle various missing data situations for full data utilization. Transformer (Vaswani et al., 2017) is currently the most versatile universal architecture that can handle any input feature type for different tasks (Carion et al., 2020; Chen et al., 2021; Dosovitskiy et al., 2020; Hatamizadeh et al., 2021; Wang et al., 2020). In the present work, transformers can address the issues associated with the multi-modal classification and with some careful design, the problem of missing data. However, applying transformers directly to just 3D medical imaging data is already extremely computationally expensive without even adding other modalities. Studies have resorted to making predictions based on 2D slices, however, 2D images are limited in the extent of the spatial coverage of the brain on medical imaging scans such as brain tumors, stroke lesions, or traumatic brain injury lesions. To utilize multi-modal information from various forms of data, a method that can take advantage of both the versatility of transformers and the rich contextual information from 3D imaging data is required.

In this paper, we propose a novel multi-modal transformer-based classification network, Multi-Modal Mixing Transformer (3MT), which is capable of automatically handling incomplete multi-modality data scenarios. This model uses CNN to efficiently extract the local feature related to AD and a transformer to capture the long-range relationship within a 3D MRI from imaging modality while cascaded multi-modal transformers are designed to aggregate imaging information with clinical information to achieve better performance. The multi-modal data in this paper include T1-weighted MRI as imaging data, as well as different clinical data such as cognitive assessments and demographic information. The model is tested to classify AD and CN patients and predict the mild cognitive impairment (MCI) conversion using multi-modal data. Our contribution is summarized as follows:

- 3MT is the first transformer-based network that is versatile enough to handle many missing data scenarios with a single model instance. Our work enables full data utilization as the network still produces the best-effort predictions based on the available modalities.
- It is an end-to-end transformer classification network that incorporates multi-modality data, including 3D MRI and 12 clinical data including demographics, cognitive test results, and genetic information, for informed disease classification. The theoretical number of modalities and feature types is unlimited.
- We propose a novel Modality Dropout (MDrop) module with Cascaded Modality Transformers (CMTs) and auxiliary output to ensure the unbiased contribution of each modality by simulating various missing data scenarios. The result is an unprecedented level of robustness to missing data.

To accurately classify, the model refines a learned query repeatedly through cascaded modality injection transformers before the final decision is made. MDrop and auxiliary prediction heads are designed to mitigate the imbalanced training of each modality. 3MT was first trained and tested on the Alzheimer's Disease Neuroimaging Initiative (ADNI) dataset without missing data for AD classification and MCI conversion prediction tasks and achieved the state-of-the-art accuracy performance. Then the model was tested given different combinations of modality missing situations on both the ADNI dataset and the The Australian Imaging, Biomarker & Lifestyle Flagship Study of Ageing (AIBL) dataset without any fine-tuning or retraining. We demonstrate the model's superior ability to handle missing data while maintaining state-of-the-art (SOTA) performance.

## 2. Related work

### 2.1. Image-based AD Detection with Deep Learning

Recently, deep neural networks have been widely applied for AD detection (Ebrahimighahnavieh et al., 2020; Wen et al., 2020). MRIs are commonly available and able to capture disease pathological patterns (Noor et al., 2019). Therefore, MRI data acquired as part of the ADNI and AIBL, which contain a fairly large number of data, have been utilized to train deep models on detecting AD pathology. Valliani and Soni (2017) fine-tuned ResNet-50 (He et al., 2016) to classify AD from CN participants using 2D MRI slices obtained from the median axial view. Qiu et al. (2018) also applied VGG-11 (Simonyan and Zisserman, 2014) to 2D MRI slices to classify MCI from CN patients. However, due to the limited coverage of the disease pattern in 2D MRI slices, deep models applied to 3D MRI data are much preferable. Lian et al. (2018) proposed a hierarchical Fully Convolutional Network (FCN) that can learn multi-scale features from small patches and whole brain regions to perform AD diagnosis and achieved 90% accuracy in classifying AD from CN. Bäckström et al. (2018) applied 3D ConvNet to study to impact of data partitioning and reported 90% accuracy on patient-level data split while 98% accuracy by a random data split. Even though there are various deep learning methods to classify AD or predict MCI outcomes, Wen et al. (2020) undertook a comprehensive review of deep models detecting AD and reported almost 50% of the published models had data leakage. Most of them are due to the wrong data split meaning that the same patient scans could appear in both the training set and validation set. Bäckström et al. (2018) also reported a 10% difference between random data spitting and patient-level data splitting. Therefore, in our work, we carefully split the ADNI dataset into training, validation, and test sets based on the patient level to ensure there was no data leakage. We also applied the AIBL dataset to test the model's performance separately.

### 2.2. Multi-modal AD Detection with Deep Learning

Even though these 2D/3D deep models using one modality have comparable results, most of the multi-modality models achieved better re-sults. Qiu et al. (2020) proposed a multi-modality FCN with multilayer perceptron (MLP) model that takes in MRIs and clinical data (age, gender, MMSE) trained on ADNI and tested on multiple datasets including AIBL, National Alzheimer's Coordinating Center (NACC) and the Framingham Heart Study (FHS). Guan et al. (2021) managed to distil information from multi-modal data to improve the MCI prediction results from MRI based prediction. Pan et al. (2020) introduced a spatially-constrained Fisher representation network that handles missing PET images according to their paired MRIs. They used a Generative Adversarial Networks (GAN) to generate missing PET images given MRIs and applied these paired data to perform brain disease detection. However, the source of the truth for the input was the same. Although multi-modal methods have yielded promising AD classification results or MCI conversion prediction results, there has yet to be a way to fully utilize existing modalities in missing data scenarios. Most of the current multi-modal methods discard samples with missing data, which causes significant data under-utilization. There are some methods that attempted to generate missing imaging data. For example, Pan et al. (2020) used a GAN to generate missing PET images. However, there has not been an effective and valid method to generate other types of missing data. Therefore, an approach that can handle missing data efficiently without introducing any bias still needs to be designed.

### 2.3. Transformers

Transformers are a type of general neural architecture that can process any data type. They were initially introduced by Vaswani et al. (2017) for machine translation tasks and achieved SOTA performance in different Natural Language Processing (NLP) tasks (Brown et al., 2020; Devlin et al., 2018; Radford et al., 2019). More recently, transformers are also adapted to computer vision tasks for image classification (Dosovitskiy et al., 2020; Jaegle et al., 2021), object detection (Carion et al., 2020) and image segmentation (Chen et al., 2021; Zheng et al., 2021). One major advantage of transformers is that it is free of inductive biases meaning they can be used to process any feature type. However, the limitation is the use of self-attention which scales poorly to image data: computational cost increases quadratically when the number of pixels increases. This is because self-attention models dense relationships between all the pixels of an image. As a result, many vision transformers are limited in input image size due to hardware limitations (Dosovitskiy et al., 2020).

To make use of the global attention mechanism while mitigating the drawbacks of transformers, plenty of works have applied CNN to extract compact spatial information, and then the size-reduced feature maps are fed into a transformer network to learn global dependencies. For example, Carion et al. (2020) applied a ResNet-based (He et al., 2016) CNN backbone to learn the useful features from the input images. Then, these features enter a downstream transformer as patch representations for further processing. Despite its simplistic and end-to-end design, DETR achieved the SOTA performance on the COCO dataset (Lin et al., 2014) for object detection tasks. For medical image segmentation, TransUNet (Chen et al., 2021) was the first to combine a CNN with a transformer for improved segmentation results. Similar to the original Unet, TransUNet applied a CNN to extract high-level spatial information, and then, transformer attention layers are used to model global dependencies. Finally, the output of the transformer is upsampled via a decoder CNN to produce a full-resolution segmentation map. Shortcuts are used to pass features from the encoder layers directly to the decoder layers to mitigate information loss. In another example, Medical transformer (MedT) (Valanarasu et al., 2021) applied gated axial-attention layers modified from Wang et al. (2020) with a Local-Global training manner to perform the medical segmentation and achieved comparable results. In AD classification task, Kushol et al. (2022) effectively utilized the frequency and image domain features from coronal 2D slices with a vision transformer architecture to classify AD and achieved SOTA performance. They have shown the global and local context and spatial features can be captured by two transformers from the image and frequency domains, respectively. We found that although all these architectures yelled SOTA results on image processing tasks, most of them were limited to 2D images/MRIs. For 3D medical image analysis where the inputs can be extremely large, a more computationally efficient is required.

The more recent Perceiver (Jaegle et al., 2021) is a type of transformer that uses cross-attention over a low-dimensional array so that the computational complexity can be reduced from quadratic to linear compared to the original design of the transformer. Perceiver's forward pass is iterative similar to Recurrent neural networks (RNNs) (Cho et al., 2014), where the input is progressively processed by the same transformer layers several times. During each iteration, additional contextual information is injected into the latent array using cross-attentions. In this way, the latent array is progressively refined with additional information and the output prediction is more accurate. Since the query of the cross-attention is a latent array (denoted $N$) that only represents class predictions, it is a constant with a relatively small latent size. Therefore, the cross-attention computation complexity is only $O(M \times N)$ where M is the size of the contextual inputs. With this property, the contextual inputs for the Perceiver are not limited by size, they can be much larger than text representations like audio, image, and point cloud inputs.

## 3. Methods

In this section, we first describe the overall architecture design of 3MT. Then we will explain in detail about different modality embeddings, modality aggregation, missing data handling using modality dropout, and the auxiliary outputs for balanced training signals. Next, the implementation details and data preprocessing steps are listed.

### 3.1. Architecture overview

The design of 3MT's architecture shown in Figure 1 is a sequence of modality transformers each incorporating features from a specific modality, and at the end, a more informed class prediction that is aggregated with extracted multi-modal features can be obtained. In this case, the class prediction starts with a learned query vector $\mathbf{Q_L}$ (512-D), which then passes through a sequence of CMTs with the injection of multi-modal information to arrive at the output AD/CN or progressive MCI (pMCI) / stable MCI (sMCI). Each CMT uses a cross-attention to inject a designated modality into $\mathbf{Q_L}$. Before being injected via CMTs, each modality is transformed into vector embeddings $\mathbf{e}$ (modality embedding) from its raw data type. In missing data scenarios, the CMTs corresponds to the missing modalities and receives zero embeddings controlled by MDrops with a certain probability for the injection, which signifies "not available" to the model. In this way, the model is trained with prior knowledge of handling all missing data scenarios. There is no theoretical limit to the number of modalities or CMTs in the sequence for 3MT to perform classification or prediction.

### 3.2. Modality embeddings

The following three types of embedding techniques are used in this case to encode three types of inputs: 1). **Categorical embedding**: a learned mapping from the raw input to a vector embedding through a lookup table. A 1-D input goes into the embedding layer and the output is a 512-D embedding feature to match the transformer dimension. This type of embedding is suitable for non-ordinal or categorical data such as gender and Apolipoprotein (APOE)4 genotyping. 2). **Ordinal embedding**: it is a learned linear transformation of the raw input to a vector through a fully connected layer. The input is also a 1-D vector and the output is a 512-D vector after linear transformation. Suitable for ordinal inputs where ordering carries significant meanings such as age and cognitive test scores. 3). **Image embedding**: it is a learned

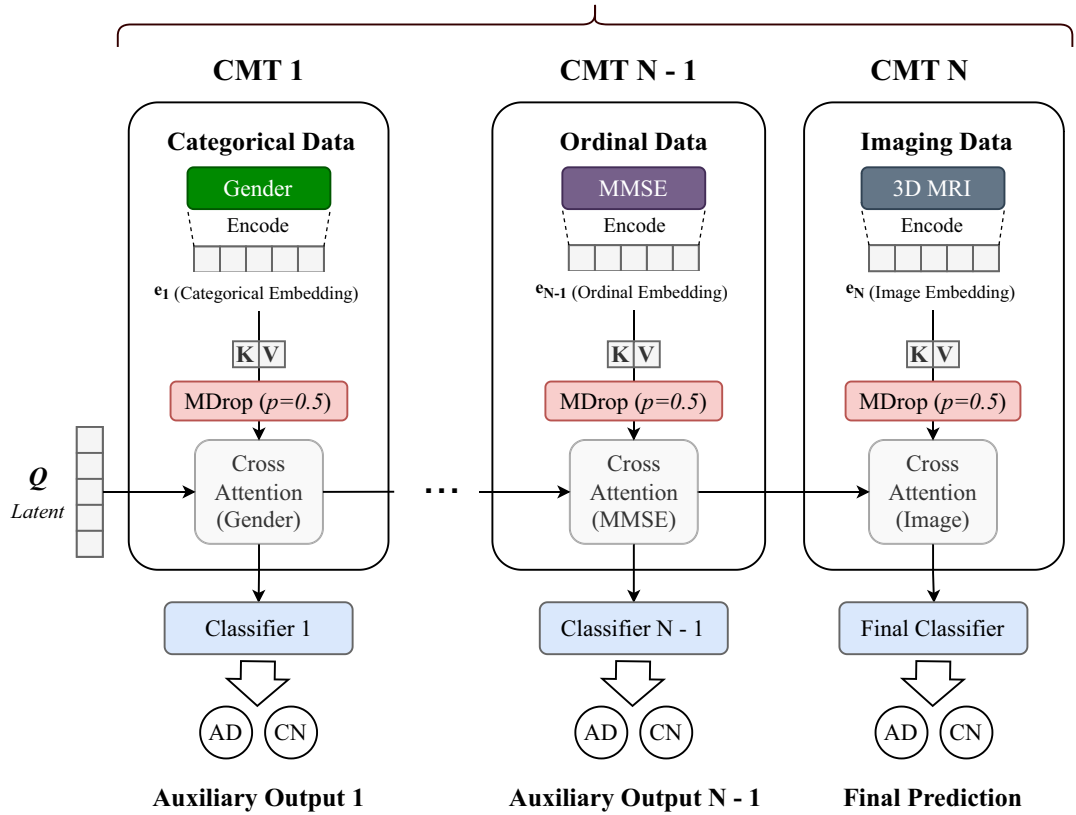

**Fig. 1.** Network architecture of 3MT. The network starts with a learned latent query vector **Q**. N multi-modal data are injected into the model via N CMTs. In each CMT, the input data is first encoded into key vectors **K** and value vectors **V**, then, the **K** and **V** vectors are injected into **Q** using cross attention. Different modalities are encoded differently according to the data type. A class prediction is generated after all the multi-modal data have been injected.

CNN-transformer-based image feature extractor that can efficiently extract local AD related features and global long-range spatial features. Details of the image feature extractor are shown in Figure 2. The image feature extractor contains four CNN blocks to consecutively downsampling the 3D input image to extract the 3D representation feature related to AD. The feature maps from the last CNN block are treated as patch representations of the input, meaning that each voxel in the feature map represents a large area in the image space. Then, these patches are embedded into a positional embedding layer to preserve the 3D spatial information before feeding into a transformer encoder to learn patch-wise correlations. Finally, a global average pooling layer and a linear projection layer are used to reduce the transformer output into a 512-D vector so that the imaging feature can be injected into the cross-attention module to update the learned input query. At the end of the last CMT in the sequence, a classifier (shallow densely-connected neural network) is used to map the transformed **Q** into two logits representing AD and CN.

### 3.3. Modality injection with CMT

CMT is a sequence of transformer decoder blocks that takes a query input $\mathbf{Q_{in}}$ from the output of the previous modality transformer or from $\mathbf{Q_L}$ if it is the first CMT in the sequence and cross-attend the extracted multi-modal features to iteratively update the learned query with more information added. The CMT consists of a multi-head self-attention layer and a multi-head cross-attention layer. Details can be found in Figure 3. In the multi-head self-attention layer, queries, keys, and values can be defined as **Q**, **K**, and **V**, respectively. Apply them h times to perform learnable linear projections. Then the self-attention dot product matrix (Vaswani et al., 2017) can be defined as follow:

$$\text{Attention}(\mathbf{Q}, \mathbf{K}, \mathbf{V}) = \text{softmax}\left(\frac{\mathbf{QK}^T}{\sqrt{d_k}}\right)\mathbf{V}. \tag{1}$$

All heads concatenated together can be calculated as follow:

$$\text{head}_i = \text{Attention}\left(\mathbf{QW}_i^Q, \mathbf{KW}_i^K, \mathbf{V}_i^V\right), \tag{2}$$

$$\text{MultiHead}(\mathbf{Q}, \mathbf{K}, \mathbf{V}) = \text{concat}(\text{head}_1, \, \dots, \, \text{head}_h)\mathbf{W}^O, \tag{3}$$

where $d_k = 512$ and $h = 8$. The cross-attention layer is similar to the self-attention layer, instead of calculating the attention to itself, cross-attention calculates the attention between inputs classes. Modality embeddings **e** are injected into $\mathbf{Q_{in}}$ using cross-attention with its key **K** and value **V**. Since transformers retain the shape of the query throughout, multiple CMTs can be cascaded to inject mixing multi-modal information into the initial query. The use of CMTs ensures that each input modality matches with a CMT and the final output is the refined result after seeing all the inputs. With this design, even if there are missing modalities, the model can still work according to the available modalities.

### 3.4. Modality dropout during training

To explicitly simulate various missing-data situations during training, we propose a novel MDrop module in each CMT to randomly zero out the corresponding modality embeddings with probability $p_{mdrop}$. The values for $p_{mdrop}$ are parameters that can be set on a per-modality basis and the default value is 0.5. Depending on the severity of the missing data situations, the value for $p_{mdrop}$ could vary. The motivation of the MDrop module is to dynamically simulate missing data scenarios during training so that the model can have prior knowledge of the missing data and learns to be more adaptive using the remaining modalities.

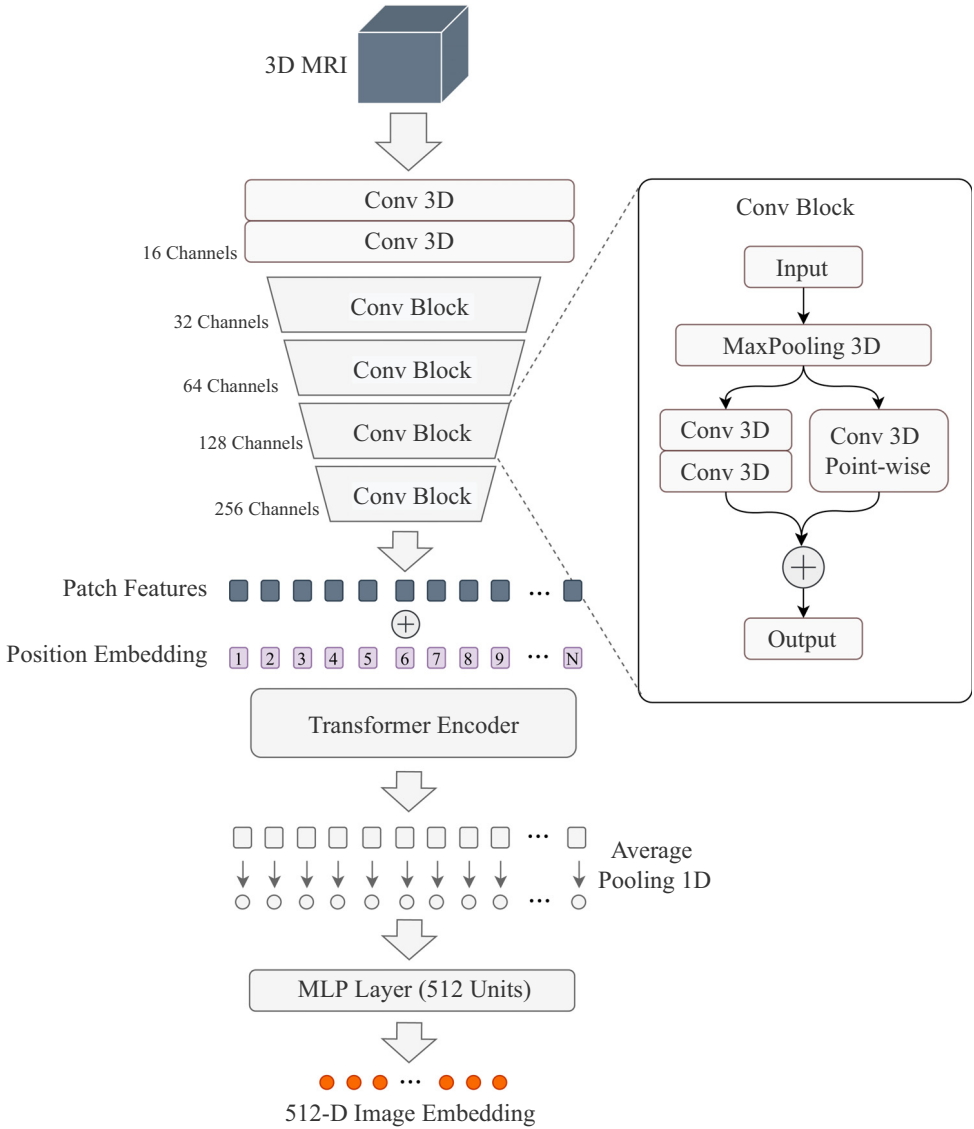

**Fig. 2.** Details of the CNN transformer for encoding 3D images. It starts with two initial convolutional layers followed by 4 residual CNN down-sampling blocks. Then, a transformer encoder is used to further learn patch-wise correlations. The output of the transformer is projected to a 512-D vector using patch-wise 1-D average pooling followed by a linear projection layer.

### 3.5. Auxiliary outputs for balanced learning signals

The sequence arrangement of CMTs can be prone to imbalanced training as the gradient struggles to flow into earlier CMTs. In response, an auxiliary classifier is added to the output of each CMT to ensure the equal contribution of the modalities. The idea is inspired by the GoogLeNet (Szegedy et al. 2015), which has auxiliary outputs connected to intermediate layers to classify the input image. While in our case, auxiliary outputs are added after adding modalities to give an intermediate output which loss will be added to the final loss. These auxiliary outputs are discarded during the inference time. Each classifier is made up of 2 densely connected layers with input and output 512 dimensions and 2 output neurons, respectively. The first densely connected layer has a LeakyReLu activation function after it. There is no weight sharing between the classifiers.

## 4. Experiments and results

### 4.1. Data preprocessing

We conducted two experiments to test 3MT performance. The first experiment is an AD classification task on the ADNI-1, ADNI-2 and ADNI-3[2] datasets (Jack et al., 2008) and the evaluation was conducted on the AIBL dataset[3] (Ellis et al., 2009). The second experiment involves predicting the conversion from MCI to pMCI or sMCI from the baseline. It is more challenging than the AD classification task, hence the results are more indicative of 3MT's classification performance. The datasets used in this experiment are ADNI-1, ADNI-2. The ADNI-3 was not included in MCI conversion prediction task since there are not enough flow-ups. Only baseline data are used for this task and there is no same patient appearing in both the ADNI-1 and the ADNI-2. The demographic information of ADNI and AIBL is shown in Table 1.

For the AD classification task, we processed overall 3194 MRIs from 816 patients. All the available time points for each patient who under-took the MPRAGE sequence were included. Repeat MRIs and other MRIs sequences were discarded. All T1-weighted MRIs were automatically reoriented to standard space and cropped using FSL. Then we applied Nonlinear registration to register MRIs to the MNI152 template. Then skull stripping and bias field correction was applied. Finally, all MRIs were cropped into 144 x 176 x 144. All available time points were included and the data split was based on the patient-level split to avoid data leakage. There are 12 types of clinical data included in the experiments: age, gender, education years, APOE4 genotyping, Clinical Dementia Rating Sum of Boxes (CDRSB), Alzheimer's disease assessment scale (ADAS) (ADAS11, ADAS13), MMSE score, and Rey Auditory Verbal Learning Test (RAVLT) (immediate, learning, forgetting, percent forgetting). All clinical data were matched with the closest MRI acquired date.

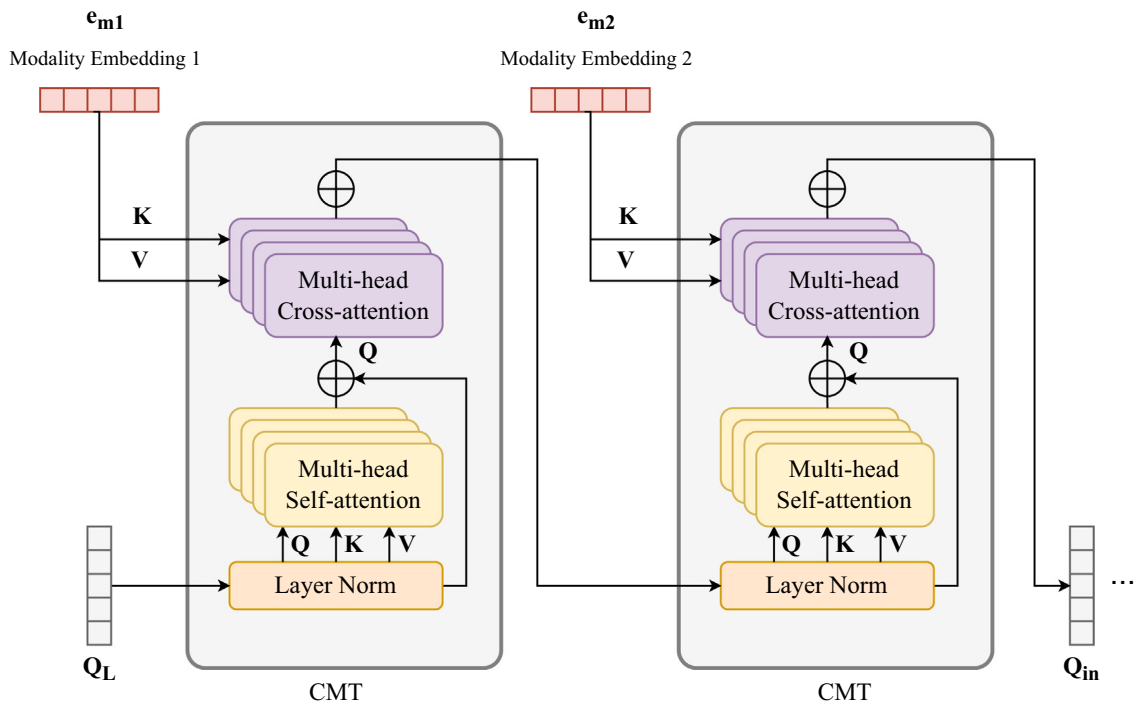

**Fig. 3.** Details of modality injection with CMT. A learned query passes through Layer Normalization then a multi-head self-attention layer and a multi-head cross-attention layer.

**Table 1**
Demographic information of study subjects of three datasets for the AD classification task.

| Dataset | Category | No. of subjects | Age Range | Female/Male | MMSE |
|---|---|---|---|---|---|
| CN | ADNI-1 | 218 | 59.9-89.6 | 107/111 | 7-30 |
| | ADNI-2 | 152 | 56.2-85.6 | 79/73 | 27-30 |
| | ADNI-3 | 149 | 55.8-86.3 | 35/18 | 24-30 |
| | AIBL | 320 | 60.0-92.0 | 176/144 | 26-30 |
| AD | ADNI-1 | 166 | 55.1-90.9 | 79/87 | 5-29 |
| | ADNI-2 | 104 | 55.9-88.5 | 45/59 | 18-27 |
| | ADNI-3 | 27 | 55.9-89.2 | 10/17 | 11-27 |
| | AIBL | 62 | 55.0-93.0 | 38/24 | 6-28 |

**Table 2**
Demographic information of study subjects of ADNI-1 and ADNI-2 for the MCI conversion prediction task.

| Dataset | Label | No. of Subjects | Age Range | Male/Female |
|---|---|---|---|---|
| ADNI1 | pMCI | 137 | 55.2-88.3 | 81/56 |
| | sMCI | 100 | 57.8-87.9 | 60/40 |
| | AD | 169 | 55.1-90.1 | 87/82 |
| | CN | 206 | 59.9-89.6 | 103/103 |
| ADNI2 | pMCI | 57 | 55.0-84.6 | 30/27 |
| | sMCI | 117 | 55.9-91.3 | 65/52 |
| | AD | 102 | 55.9-88.3 | 58/44 |
| | CN | 137 | 56.2-85.6 | 67/70 |

For the MCI conversion prediction task, we processed patients only at the baseline from AD, CN, and MCI groups from the ADNI1 and ADNI2 datasets. We acknowledge the potential bias in deep learning prediction of pMCI and sMCI due to variations in a magnetic resonance field strength Thibeau-Sutre et al. (2022). Therefore, we ensured that all 612 patients from the ADNI1 dataset were scanned using 1.5T T1-weighted MRI images, while all 413 patients from the ADNI2 dataset were scanned using 3T T1-weighted MRI images. The ADNI1 and ADNI2 datasets were used for training and testing separately, and subsequently, the training and testing sets were swapped to eliminate any fixed 1.5T or 3T MRI biases in either set. The AD and the CN patients are diagnosed as AD or CN at all available points and did not get better or worse. The MCI group can be divided into two groups: sMCI and pMCI. The sMCI group can be defined as the patient being diagnosed as MCI at all available time points with at least 3 years of diagnosis records. The pMCI group was defined as patients who were diagnosed as MCI at the baseline while converted to AD in 3 years and also not reported to convert back to MCI or CN at all the available follow-ups. Other patients not meet these criteria were excluded. We applied the same data preprocessing steps as the AD classification task and collected the same 12 clinical data, which were matched with the closest to the baseline timepoint. Details of the demographics of patients are shown in Table 2.

### 4.2. Experimental setup

3MT was implemented in Pytorch[4] and trained on an Nvidia V100 GPU with 32GB of memory. The model is trained for 50 epochs using the Adam optimizer (Kingma and Ba, 2014) with a batch size of 2. The learning rate of all experiments was set to be $5e^{-4}$ and the cross-entropy loss was applied as the loss function. Random coronal-view flipping and Gaussian noising were applied as data augmentation. In the MCI conversion prediction experiment, patients diagnosed as AD and CN at the baseline were also applied as the training data augmentation. AD and pMCI are labelled as 1 while CN and sMCI are labelled as 0. The query and transformers dimension are 512 while each transformer layer has 8 heads. One of the key advantages of 3MT is the automatic aggregation of different modalities to produce the final prediction meaning that 3MT can infer given one or any mixture of modalities after training. We first show that 3MT can achieve the SOTA performance on classifying AD using imaging data with clinical data on the ADNI dataset. We applied five-fold cross-validation to the ADNI dataset and then further tested the model performance on the AIBL dataset with missing data. This is due to the AIBL dataset does not provide as much clinical data as the ADNI dataset, which provided us an opportunity to evaluate our model in missing data situations. Only MRI, age, gender and MMSE were applied during testing.

---
[4] https://pytorch.org/

On the other hand, we also trained 3MT on a harder task, to predict the MCI conversion using only the ADNI dataset. AIBL dataset was not applied because it does not provide enough MCI patients. We first trained the model on the ADNI-1 dataset and evaluated using the ADNI-2 dataset. Then we swap training and testing datasets meaning that we trained on the ADNI-2 dataset and evaluated on the ADNI-1 dataset. The train and validation ratio is 8:2 while the highest AUC value on the validation set is selected as the final weight and applied to the test set. Each experiment repeats five times and took the average. All the networks are trained with the complete data (MRI + clinical data) with/without MDrop module and missing data scenarios are tested during model evaluation.

### 4.3. Evaluation metrics

The metrics used to evaluate model performance were accuracy, sensitivity, specificity, and area under curve (AUC) score. AUC score is chosen since it is a better metric than accuracy to measure the model performance on imbalanced data. The problem with accuracy is that it could hide the true performance with high specificity and low sensitivity and vice versa. The metrics are calculated as:

$$Accuracy = (TP + TN) / (TP + TN + FN + FP) \qquad (4)$$

$$Sensitivity = TP / (FN + TP) \qquad (5)$$

$$Specificity = TN / (TN + FP) \qquad (6)$$

where TP, TN, FN, and FP are true positive, true negative, false positive, and false negative, respectively. In table 5, we show the model performance using these four metrics on the ADNI test set and the AIBL dataset.

### 4.4. Classification results with complete data

Table 3 compares 3MT with other SOTA AD classification models without missing data. Many methods in the literature are limited to MRI only, and their results have been included in the top section of the table for reference. The bottom section of the table compares the performance of 3MT with previous works that also used multi-modal data. As previously described, methods with data leakage are not included in this comparison. Compare with the only MRI as the single modality input, our multi-modal network has significantly better accuracy and AUC results. Some previous studies also applied transformer or CNN with transformer (Jang and Hwang, 2022; Kushol et al., 2022; Li et al., 2022), our 3MT model with multi-modal data increases at least 6% of accuracy and 3% of AUC. Compare with other multi-modal works, our framework outperforms (Pan et al., 2020), which used multi-modal medical image data including PET and MRI by 5.8% and 2.7% in terms of accuracy and AUC, respectively. Qiu et al. (2020) also used MRI and clinical data and our framework achieved better accuracy by 2.6%. It is evident that 3MT is versatile enough to aggregate knowledge from diverse imaging and clinical data to produce superior AD classification results. Note that due to the different number of subjects and data pre-processing methods, direct comparison between different methods may be affected. We provided the number of subjects for each study and we collected the most number of AD and CN subjects for the AD classification task.

The MCI conversion prediction results are presented in Table 4. Here, we compare 3MT with other SOTA methods without missing data. To ensure a fair comparison, we followed most of the literature that used the 36 months time point to determine sMCI or pMCI patients. Our multi-modal framework outperforms the MRIs-only work (Jang and Hwang, 2022) by a significant margin, at the same time, it also achieved the best performance compared to other multi-modal methods when trained on the ADNI-1 dataset and evaluated it on the ADNI-2 dataset. To compare with Guan et al. (2021), we applied the same data preprocessing steps, our model's AUC is 3.17% and 2.79% better on the whole ADNI-1 and ADNI-2 as test set, respectively.

The SOTA results of 3MT without missing data from these two tasks suggest that is a suitable architecture for multi-modal classification or prediction, and it is also versatile enough to handle vastly diverse feature types.

### 4.5. Classification results with missing data

To demonstrate the performance of 3MT when facing missing data scenarios in the real world, we evaluated it on the AIBL dataset which lacks clinical data such as cognitive tests. Only 4 age, gender, APOE4, and MMSE were used as the clinical data input for this evaluation. For training, the model uses the ADNI dataset with all 12 clinical data, and for evaluation on AIBL we test 1). Only providing MRIs when all the clinical data are missing, 2). only providing 4 basic clinical data when imaging data and the other 8 advanced clinical data are missing, 3). providing both imaging data and 4 basic clinical data when the other 8 advanced clinical data are missing. In this case, we applied the same AIBL test set as Qiu et al. (2020) using the same clinical data to be able to have a fair comparison. As shown in Table 5, when there is only MRI or clinical data provided using an external AIBL dataset, our framework has a 3.3% and 0.9% accuracy improvement compared to Qiu et al. (2020), respectively. When only 4 clinical data with MRI are provided, our model improved accuracy from 0.932 to 0.963. Note that our model is trained with full data and evaluated with only MRI, clinical data or MRI with 4 clinical data and 8 missing clinical data while Qiu et al. (2020)'s model was trained and evaluated with the full data. We demonstrated that even with some important missing data, 3MT can still achieve reasonably good results on different datasets.

### 4.6. Ablation Study

We conducted ablation studies on each of 3MT's modules. These models were independently removed and the classification performance was accessed. The experiments here include 1). 3MT without the image transformer and using MRI the only input modality, 2) 3MT without MDrop and using data with missing information, 3), 3MT without the Auxiliary outputs. The results are shown in Table 6. In the top half of the table, we demonstrated that adding the patch-based image transformer block can increase the model's performance on both ADNI and AIBL datasets when only feeding in the MRI data. This shows that adding a transformer block can better capture the long-range correlation between patches. In the bottom half of the table, we first show when the MDrop block is removed, the model classification AUC with missing data dropped by 23% and 34% on the ADNI dataset and AIBL dataset, respectively. This means that the MDrop block plays an important role

**Table 3**

AD classification comparison between 3MT and SOTA methods. The first half of the table shows the performance of SOTA methods using MRI as a single modality input. The second half of the table shows the performance of SOTA methods and 3MT using multi-modal data.

| Study | Methods | No. of subjects | Modalities | Accuracy | Sensitivity | Specificity | AUC |
|---|---|---|---|---|---|---|---|
| Valliani and Soni (2017) | ResNet-50 | 660 | MRI | 0.810 | - | - | - |
| Lian et al. (2018) | Hierarchical FCN | 787 | MRI | 0.900 | 0.820 | 0.970 | - |
| Wen et al. (2020) | CNN +FCN | 666 | MRI | 0.850 | - | - | - |
| Jang and Hwang (2022) | CNN+Transformer | 751 | MRI | 0.932 | - | - | 0.963 |
| Pan et al. (2020) | Spatially-constrained Fisher representation | 810 | MRI | 0.914 | 0.897 | 0.928 | 0.962 |
| Li et al. (2022) | Transformer | 533 | MRI | 0.939 | 0.905 | 0.957 | 0.968 |
| Kushol et al. (2022) | Tranformer | 388 | MRI | 0.882 | 0.956 | 0.774 | - |
| Pan et al. (2020) | Spatially-constrained Fisher representation | 810 | MRI+PET | 0.936 | 0.915 | 0.952 | 0.970 |
| Qiu et al. (2020) | FCN+MLP | 487 | MRI+Clinical data | 0.968 | 0.957 | 0.977 | - |
| **Our model** | CNN+Transformer + CMT | 816 | MRI+Clinical data | 0.994 | 1.000 | 0.989 | 0.997 |

**Table 4**

Quantitative comparisons between 3MT and SOTA methods for predicting MCI to AD conversion on ADNI1 and ADNI2.

| Model | No. of MCI | Modalities | ADNI1 | | ADNI2 | |
|---|---|---|---|---|---|---|
| | | | AUC | ACC | AUC | ACC |
| Jang and Hwang (2022) | 411 | MRI | 66.45 | 62.45 | 73.56 | 72.05 |
| Pan et al. (2020) | 694 | MRI + PET | - | - | 83.32 | 77.77 |
| Qiu et al. (2020) | 411 | MRI + Clinical data | 71.82 | 70.11 | 72.03 | 66.90 |
| Lin et al. (2018) | 308 | MRI + Clinical data | 86.10 | 79.9 | - | - |
| Guan et al. (2021) | 455 | MRI + Clinical data | 80.80 | 74.60 | 87.10 | 80.00 |
| Our model | 411 | MRI + Clinical data | 83.97 | 76.73 | 89.89 | 83.33 |

**Table 5**

Classification results with missing data on ADNI test set and AIBL dataset.

| Methods | Modality | Testset | Accuracy | Sensitivity | Specificity | AUC |
|---|---|---|---|---|---|---|
| 3MT (With Dropout) | MRI | ADNI | 0.914 | 0.954 | 0.864 | 0.957 |
| | Clinical data | ADNI | 0.918 | 0.945 | 0.892 | 0.954 |
| | MRI + Clinical data | ADNI | 0.930 | 0.956 | 0.903 | 0.970 |
| | MRI | AIBL | 0.903 | 0.928 | 0.806 | 0.916 |
| | Clinical data | AIBL | 0.924 | 0.937 | 0.855 | 0.978 |
| | MRI + Clinical data | AIBL | **0.963** | **0.975** | 0.903 | **0.984** |
| Qiu et al. (2020) | MRI | AIBL | 0.870 | 0.594 | 0.924 | - |
| | Clinical data | AIBL | 0.915 | 0.872 | 0.923 | - |
| | MRI + Clinical data | AIBL | 0.932 | 0.877 | **0.943** | - |

**Table 6**

Quantitative comparison of AD classification using 5 different variants of 3MT to evaluate the effect of each module on the performance.

| Model | ADNI | | AIBL | |
|---|---|---|---|---|
| | AUC | ACC | AUC | ACC |
| w/o Transformer block | 0.926 | 0.888 | 0.895 | 0.882 |
| 3MT(Image only) | 0.957 | 0.914 | 0.916 | 0.903 |
| w/o Dropout | 0.742 | 0.696 | 0.650 | 0.770 |
| w/o Auxiliary output | 0.965 | 0.891 | 0.983 | 0.926 |
| 3MT | **0.970** | **0.930** | **0.984** | **0.963** |

during training for the model to adapt to different missing data scenarios. Therefore, when there are missing modalities without the MDrop block, the performance decreases dramatically. We further show that adding the Auxiliary output block can further increase the model performance in both AUC and ACC metrics, which means that only passing the learned query information is not enough for the model to mitigate the imbalanced signal during training from multi-modal data.

### 4.7. Effect of modality dropout

In this section, we show that the model performance change according to the different $p_{mdrop}$ values in both full data and missing data scenarios. The $p_{mdrop}$ value during the training process starts from 0.9, meaning 90% of chance each modality was dropped, until 0, meaning that no modality was dropped. In Figure 4, full data including 12 clinical data and 3D MRI are provided during evaluation, when $p_{mdrop}$ decreases, the AD classification accuracy increases. Therefore, $p_{mdrop}$ is set to be 0 when there is no missing modality, 3MT can achieve the best performance. In Figure 5, when the model is provided with all missing cognitive test results, without the MDrop module results in a severe performance drop. When $p_{mdrop} = 0.5$ is the optimal hyperparameter for AD classification, other $p_{mdrop}$ value also plays an important role in handling missing data situations. $p_{mdrop}$ value may vary according to the severity of missing data in different datasets.

## 5. Limitations and future work

There are several limitations to the current form of 3MT that should be acknowledged. Firstly, while 3MT is memory efficient compared to other transformer and 3D CNN models, its reliance on a GPU may pose challenges in clinical environments where access to GPUs is limited. It would be important to explore strategies for deploying 3MT on hardware that is more readily available in clinical settings. Secondly, we did not conduct the MCI prediction experiment using incomplete data due to the lack of existing literature on the same task for comparison. Furthermore, the AIBL dataset, as mentioned by Guan et al. (2021), does not provide a sufficient number of MCI patients and clinical data. In future work, it would be valuable to test the performance of our model on the MCI prediction task using a suitable dataset if one becomes available. Another limitation pertains to the potential feature leakage when combining clinical and imaging data for AD vs CN classification with complete data. In practice, clinicians often employ similar features to differentiate AD from CN individuals, which could introduce biases. While our model was trained using complete data with 12 clinical features, during testing we considered various missing data scenarios and utilized only four basic clinical features (age, gender, APOE4, and MMSE score) to ensure a fair comparison with previous work (Qiu et al., 2020). We deliberately avoided incorporating advanced cognitive tests during evaluation. Nonetheless, future research should explore methods to mitigate feature leakage and develop strategies that better align with clinical practices. For future work, an alternative to clinical diagnosis, involving biomarker-based approaches such as amyloid and/or tau measurements, holds promise for improving diagnostic accuracy and reliability Ossenkoppele et al. (2022); Teunissen et al. (2022). Incorporating these measures, obtained through imaging or blood-based tests, into deep learning models could enhance their performance. Furthermore, considering multi-modal imaging data, including FDG-PET and diffusion MRI, may provide additional insights and further improve the model's diagnostic capabilities. Exploring transfer learning and self-supervision techniques is also an avenue worth pursuing. These approaches have the potential to enhance the model's capabilities by leveraging pre-training on large-scale datasets or incorporating self-supervised learning methods. By integrating self-supervised learning, the model may benefit from learning more representative and robust features, leading to improved performance in future iterations. Addressing these limitations and exploring the proposed future directions could contribute to the advancement and practical application of 3MT in the diagnosis and prediction of AD and MCI.

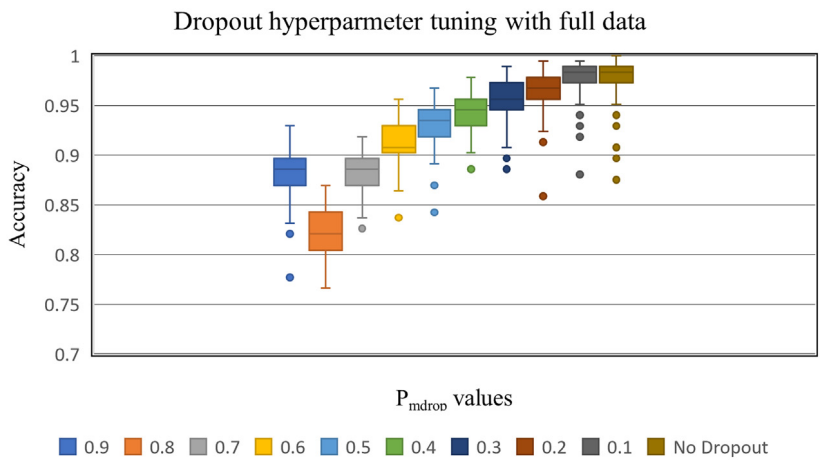

**Fig. 4.** Effects of different values of $p_{mdrop}$ (ranging between 0 and 0.9) on the classification accuracy with full data.

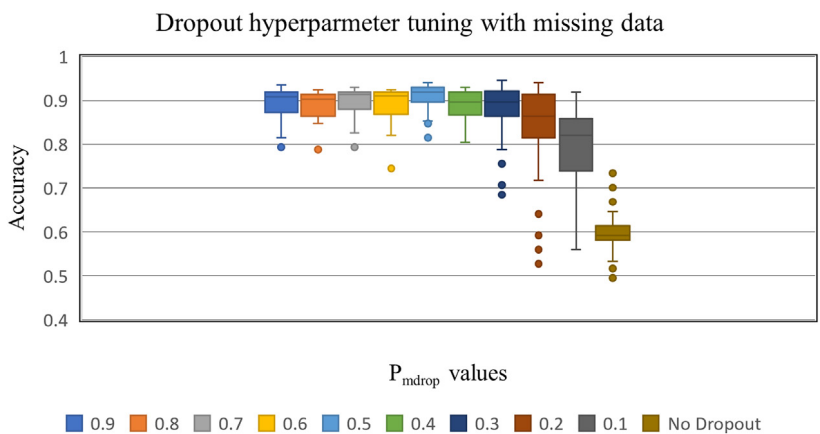

**Fig. 5.** Effects of different values of $p_{mdrop}$ (ranging between 0 and 0.9) on the classification accuracy when cognitive test data are missing.

## 6. Conclusion

In conclusion, our study highlights the performance of the proposed 3MT model in classifying AD and predicting MCI conversion using complete multi-modal data. Moreover, we have demonstrated its unprecedented robustness in AD classification tasks using incomplete data from both the ADNI and AIBL datasets. The model effectively leverages the versatility of CMTs' cross-attention mechanism to dynamically integrate various types of data. To ensure optimal performance, advanced mechanisms such as CMT mixing of transformer outputs, MDrop modules, and auxiliary outputs were incorporated during training to effectively balance the influence of each modality. This approach allows the transformer to handle missing data combinations flexibly, leading to highly reliable predictions. Additionally, we observed that the 3MT model maintains its SOTA performance even when faced with domain shifts between the ADNI and AIBL datasets. In the future, the integration of amyloid and tau measures, obtained through imaging or blood-based assessments, holds great importance for AD-related deep learning models. These biomarker-based approaches hold great promise for improving the diagnostic accuracy and reliability of AD classification. Furthermore, the exploration of additional imaging modalities and the refinement of learning techniques will contribute to the continuous advancement of AD diagnosis and research. Overall, our study demonstrates the effectiveness of the 3MT model in handling incomplete multi-modal data for AD classification across multiple datasets. As the field progresses, the integration of emerging biomarker measures and the exploration of novel imaging modalities will further enhance the capabilities of deep learning models in the AD research and diagnosis.

## Ethics statement

In accordance with the Declaration of Helsinki, all study participants provided written informed consent, following the protocols established by ADNI and AIBL. Additional details can be found at https://adni.loni.usc.edu/ and https://aibl.csiro.au/. Moreover, the data collection and sharing for this study have received ethical approval from the institutional review boards of the participating institutions in the ADNI and AIBL. Note that this paper does not involve any human participants conducted by any of the authors.

## Data and Code Availability

The Alzheimer's Disease Neuroimaging Initiative (ADNI) data and The Australian Imaging, Biomarker & Lifestyle Flagship Study of Ageing (AIBL) were used in this study. ADNI and AIBL data is publicly available and can be downloaded from https://adni.loni.usc.edu/ and https://aibl.csiro.au/, respectively. All data splits are available here: https://github.com/LinfengLiu98/3MT. Code will be available from the corresponding author upon request.

## Declaration of Competing Interest

The authors declare that they have no known competing financial interests or personal relationships that could have appeared to influence the work reported in this paper.

## Credit authorship contribution statement

**Linfeng Liu:** Conceptualization, Data curation, Writing – original draft. **Siyu Liu:** Methodology, Validation, Writing – original draft. **Lu Zhang:** Supervision, Writing – review & editing. **Fatima Nasrallah:** Supervision, Funding acquisition, Writing – review & editing. **Shekhar S. Chandra:** Supervision, Methodology, Writing – review & editing.

## Data availability

The authors do not have permission to share data.

## Acknowledgment

This research was supported by the Medical Research Future Fund (MRF1201961) and the Motor Accident Insurance Commission (MAIC), The Queensland Government, Australia (grant number: 2014000857).

Data collection and sharing for this project was funded by the Alzheimer's Disease Neuroimaging Initiative (ADNI) (National Institutes of Health Grant U01 AG024904) and DOD ADNI (Department of Defense award number W81XWH-12-2-0012). ADNI is funded by the National Institute on Aging, the National Institute of Biomedical Imaging and Bioengineering, and through generous contributions from the following: AbbVie, Alzheimer's Association; Alzheimer's Drug Discovery Foundation; Araclon Biotech; BioClinica, Inc.; Biogen; Bristol-Myers Squibb Company; CereSpir, Inc.; Cogstate; Eisai Inc.; Elan Pharmaceuticals, Inc.; Eli Lilly and Company; EuroImmun; F. Hoffmann-La Roche Ltd and its affiliated company Genentech, Inc.; Fujirebio; GE Healthcare; IXICO Ltd.;Janssen Alzheimer Immunotherapy Research & Development, LLC.; Johnson & Johnson Pharmaceutical Research & Development LLC.; Lumosity; Lundbeck; Merck & Co., Inc.;Meso Scale Diagnostics, LLC.; NeuroRx Research; Neurotrack Technologies; Novartis Pharmaceuticals Corporation; Pfizer Inc.; Piramal Imaging; Servier; Takeda Pharmaceutical Company; and Transition Therapeutics. The Canadian Institutes of Health Research is providing funds to support ADNI clinical sites in Canada. Private sector contributions are facilitated by the Foundation for the National Institutes of Health (www.fnih.org). The grantee organization is the Northern California Institute for Research and Education, and the study is coordinated by the Alzheimer's Therapeutic Research Institute at the University of Southern California. ADNI data are disseminated by the Laboratory for Neuro Imaging at the University of Southern California. The authors would also like to acknowledge The Australian Imaging Biomakers and Lifestyle (AIBL) for data collection and sharing. Details of the AIBL can be found at www.AIBL.csiro.au and a list of the researchers of AIBL is provided at https://aibl.csiro.au/.

We thank Dr. Xinwen Liu for her expertise and advice on the study.

## References

Association, A., 2019. 2019 alzheimer's disease facts and figures. Alzheimer's & dementia 15 (3), 321–387.

Beach, T.G., Monsell, S.E., Phillips, L.E., Kukull, W., 2012. Accuracy of the clinical diagnosis of alzheimer disease at national institute on aging alzheimer disease centers, 2005–2010. Journal of neuropathology and experimental neurology 71 (4), 266–273.

Brown, T.B., Mann, B., Ryder, N., Subbiah, M., Kaplan, J., Dhariwal, P., Neelakantan, A., Shyam, P., Sastry, G., Askell, A., et al., 2020. Language models are few-shot learners. arXiv preprint arXiv:2005.14165.

Bäckström, K., Nazari, M., Gu, I.Y.-H., Jakola, A.S., 2018. An efficient 3d deep convolutional network for alzheimer's disease diagnosis using MR images. In: 2018 IEEE 15th International Symposium on Biomedical Imaging (ISBI 2018), pp. 149–153. doi:10.1109/ISBI.2018.8363543.

Carion, N., Massa, F., Synnaeve, G., Usunier, N., Kirillov, A., Zagoruyko, S., 2020. End-to-end object detection with transformers. In: European Conference on Computer Vision. Springer, pp. 213–229.

Chen, J., Lu, Y., Yu, Q., Luo, X., Adeli, E., Wang, Y., Lu, L., Yuille, A.L., Zhou, Y., 2021. Transunet: Transformers make strong encoders for medical image segmentation. arXiv preprint arXiv:2102.04306.

Cho, K., Van Merriënboer, B., Gulcehre, C., Bahdanau, D., Bougares, F., Schwenk, H., Bengio, Y., 2014. Learning phrase representations using RNN encoder-decoder for statistical machine translation. arXiv preprint arXiv:1406.1078.

Devlin, J., Chang, M.-W., Lee, K., Toutanova, K., 2018. Bert: Pre-training of deep bidirectional transformers for language understanding. arXiv preprint arXiv:1810.04805.

Dosovitskiy, A., Beyer, L., Kolesnikov, A., Weissenborn, D., Zhai, X., Unterthiner, T., Dehghani, M., Minderer, M., Heigold, G., Gelly, S., et al., 2020. An image is worth 16x16 words: Transformers for image recognition at scale. arXiv preprint arXiv:2010.11929.

Ebrahimighahnavieh, M.A., Luo, S., Chiong, R., 2020. Deep learning to detect alzheimer's disease from neuroimaging: A systematic literature review. Computer methods and programs in biomedicine 187, 105242.

Ellis, K.A., Bush, A.I., Darby, D., De Fazio, D., Foster, J., Hudson, P., Lautenschlager, N.T., Lenzo, N., Martins, R.N., Maruff, P., et al., 2009. The australian imaging, biomarkers and lifestyle (AIBL) study of aging: methodology and baseline characteristics of 1112 individuals recruited for a longitudinal study of alzheimer's disease. International psychogeriatrics 21 (4), 672–687.

Guan, H., Wang, C., Tao, D., 2021. Mri-based alzheimer's disease prediction via distilling the knowledge in multi-modal data. arXiv preprint arXiv:2104.03618.

Hanseeuw, B.J., Betensky, R.A., Jacobs, H.I.L., Schultz, A.P., Sepulcre, J., Becker, J.A., Cosio, D.M.O., Farrell, M., Quiroz, Y.T., Mormino, E.C., et al., 2019. Association of amyloid and tau with cognition in preclinical alzheimer disease: a longitudinal study. JAMA neurology 76 (8), 915–924.

Hatamizadeh, A., Tang, Y., Nath, V., Yang, D., Myronenko, A., Landman, B., Roth, H., Xu, D., 2021. Unetr: Transformers for 3d medical image segmentation. arXiv preprint arXiv:2103.10504.

He, K., Zhang, X., Ren, S., Sun, J., 2016. Deep residual learning for image recognition. In: Proceedings of the IEEE conference on computer vision and pattern recognition, pp. 770–778.

Humpel, C., 2011. Identifying and validating biomarkers for alzheimer's disease. Trends in biotechnology 29 (1), 26–32.

Hutton, C., Draganski, B., Ashburner, J., Weiskopf, N., 2009. A comparison between vox-el-based cortical thickness and voxel-based morphometry in normal aging. Neuroim-age 48 (2), 371–380.

Jack Jr, C.R., Bernstein, M.A., Fox, N.C., Thompson, P., Alexander, G., Harvey, D., Borowski, B., Britson, P.J., Jennifer, L.W., Ward, C., et al., 2008. The alzheimer's dis-ease neuroimaging initiative (ADNI): MRI methods. Journal of Magnetic Resonance Imaging: An Official Journal of the International Society for Magnetic Resonance in Medicine 27 (4), 685–691.

Jaegle, A., Gimeno, F., Brock, A., Zisserman, A., Vinyals, O., Carreira, J., 2021. Perceiver: General perception with iterative attention. arXiv preprint arXiv:2103.03206.

Jang, J., Hwang, D., 2022. M3t: Three-dimensional medical image classifier using mul-ti-plane and multi-slice transformer. In: Proceedings of the IEEE/CVF Conference on Computer Vision and Pattern Recognition (CVPR), pp. 20718–20729.

Kingma, D.P., Ba, J., 2014. Adam: A method for stochastic optimization. arXiv preprint arXiv:1412.6980.

Klöppel, S., Stonnington, C.M., Chu, C., Draganski, B., Scahill, R.I., Rohrer, J.D., Fox, N.C., Jack Jr, C.R., Ashburner, J., Frackowiak, R.S.J., 2008. Automatic classification of MR scans in alzheimer's disease. Brain 131 (3), 681–689.

Kushol, R., Masoumzadeh, A., Huo, D., Kalra, S., Yang, Y.-H., 2022. Addformer: Alzheimer's disease detection from structural mri using fusion transformer. In: 2022 IEEE 19th International Symposium on Biomedical Imaging (ISBI). IEEE, pp. 1–5.

Lebedev, A.V., Westman, E., Van Westen, G., Kramberger, M.G., Lundervold, A., Aars-land, D., Soininen, H., Kłoszewska, I., Mecocci, P., Tsolaki, M., et al., 2014. Random forest ensembles for detection and prediction of alzheimer's disease with a good be-tween-cohort robustness. NeuroImage: Clinical 6, 115–125.

Li, C., Cui, Y., Luo, N., Liu, Y., Bourgeat, P., Fripp, J., Jiang, T., 2022. Trans-resnet: In-tegrating transformers and CNNs for alzheimer's disease classification. In: 2022 IEEE 19th International Symposium on Biomedical Imaging (ISBI). IEEE, pp. 1–5.

Lian, C., Liu, M., Zhang, J., Shen, D., 2018. Hierarchical fully convolutional network for joint atrophy localization and alzheimer's disease diagnosis using structural MRI. IEEE transactions on pattern analysis and machine intelligence 42 (4), 880–893.

Lin, T.-Y., Maire, M., Belongie, S., Hays, J., Perona, P., Ramanan, D., Dollár, P., Zit-nick, C.L., 2014. Microsoft coco: Common objects in context. In: European conference on computer vision. Springer, pp. 740–755.

Lin, W., et al. 2018. Convolutional neural networks-based MRI image analysis for the alzheimer's disease prediction from mild cognitive impairment. Frontiers in neuro-science 12, 777.

McKhann, G., Drachman, D., Folstein, M., Katzman, R., Price, D., Stadlan, E.M., 1984. Clin-ical diagnosis of alzheimer's disease: Report of the NINCDS-ADRDA work group* un-der the auspices of department of health and human services task force on alzheimer's disease. Neurology 34 (7). 939–939

McKhann, G.M., Knopman, D.S., Chertkow, H., Hyman, B.T., Jack Jr, C.R., Kawas, C.H., Klunk, W.E., Koroshetz, W.J., Manly, J.J., Mayeux, R., et al., 2011. The diagnosis of dementia due to alzheimer's disease: recommendations from the national institute on aging-alzheimer's association workgroups on diagnostic guidelines for alzheimer's disease. Alzheimer's & dementia 7 (3), 263–269.

Meredith Jr, J.E., Sankaranarayanan, S., Guss, V., Lanzetti, A.J., Berisha, F., Neely, R.J., Slemmon, J.R., Portelius, E., Zetterberg, H., Blennow, K., et al., 2013. Characterization of novel CSF tau and ptau biomarkers for alzheimer's disease. PloS one 8 (10), e76523.

Noor, M.B.T., Zenia, N.Z., Kaiser, M.S., Mahmud, M., Mamun, S.A., 2019. Detecting neu-rodegenerative disease from MRI: a brief review on a deep learning perspective. In: International conference on brain informatics. Springer, pp. 115–125.

Ossenkoppele, R., van der Kant, R., Hansson, O., 2022. Tau biomarkers in alzheimer's disease: towards implementation in clinical practice and trials. The Lancet Neurology.

Pan, Y., Liu, M., Lian, C., Xia, Y., Shen, D., 2020. Spatially-constrained fisher representa-tion for brain disease identification with incomplete multi-modal neuroimages. IEEE Transactions on Medical Imaging 39 (9), 2965–2975.

Qiu, S., Chang, G.H., Panagia, M., Gopal, D.M., Au, R., Kolachalama, V.B., 2018. Fusion of deep learning models of MRI scans, mini–mental state examination, and logical mem-ory test enhances diagnosis of mild cognitive impairment. Alzheimer's & Dementia: Diagnosis, Assessment & Disease Monitoring 10, 737–749.

Qiu, S., Joshi, P.S., Miller, M.I., Xue, C., Zhou, X., Karjadi, C., Chang, G.H., Joshi, A.S., Dwyer, B., Zhu, S., et al., 2020. Development and validation of an interpretable deep learning framework for alzheimer's disease classification. Brain 143 (6), 1920–1933.

Radford, A., Wu, J., Child, R., Luan, D., Amodei, D., Sutskever, I., et al., 2019. Language models are unsupervised multitask learners. OpenAI blog 1 (8), 9.

Servick, K, 2019. Another major drug candidate targeting the brain plaques of alzheimer's disease has failed. what's left. Science 10.

Simonyan, K., Zisserman, A., 2014. Very deep convolutional networks for large-scale im-age recognition. arXiv preprint arXiv:1409.1556.

Small, G.W., Kepe, V., Ercoli, L.M., Siddarth, P., Bookheimer, S.Y., Miller, K.J., Lavret-sky, H., Burggren, A.C., Cole, G.M., Vinters, H.V., et al., 2006. Pet of brain amyloid and tau in mild cognitive impairment. New England Journal of Medicine 355 (25), 2652–2663.

Szegedy, C., Liu, W., Jia, Y., Sermanet, P., Reed, S., Anguelov, D., Erhan, D., Vanhoucke, V., Rabinovich, A., 2015. Going deeper with convolutions. In: Proceedings of the IEEE conference on computer vision and pattern recognition, pp. 1–9.

Teunissen, C.E., Verberk, I.M.W., Thijssen, E.H., Vermunt, L., Hansson, O., Zetterberg, H., van der Flier, W.M., Mielke, M.M., Del Campo, M., 2022. Blood-based biomarkers for alzheimer's disease: towards clinical implementation. The Lancet Neurology 21 (1), 66–77.

Thibeau-Sutre, E., Couvy-Duchesne, B., Dormont, D., Colliot, O., Burgos, N., 2022. Mri field strength predicts alzheimer's disease: a case example of bias in the adni data set. In: 2022 IEEE 19th International Symposium on Biomedical Imaging (ISBI). IEEE, pp. 1–4.

Valanarasu, J.M.J., Oza, P., Hacihaliloglu, I., Patel, V.M., 2021. Medical trans-former: Gated axial-attention for medical image segmentation. arXiv preprint arXiv:2102.10662.

Valliani, A., Soni, A., 2017. Deep residual nets for improved alzheimer's diagnosis. In: Pro-ceedings of the 8th ACM International Conference on Bioinformatics, Computational Biology, and Health Informatics. 615–615

Vaswani, A., Shazeer, N., Parmar, N., Uszkoreit, J., Jones, L., Gomez, A.N., Kaiser, L., Polosukhin, I., 2017. Attention is all you need. In: Advances in neural information processing systems, pp. 5998–6008.

Wang, H., Zhu, Y., Green, B., Adam, H., Yuille, A., Chen, L.-C., 2020. Axial-deeplab: Stand-alone axial-attention for panoptic segmentation. In: European Conference on Computer Vision. Springer, pp. 108–126.

Wen, J., Thibeau-Sutre, E., Diaz-Melo, M., Samper-González, J., Routier, A., Bottani, S., Dormont, D., Durrleman, S., Burgos, N., Colliot, O., et al., 2020. Convolutional neural networks for classification of alzheimer's disease: Overview and reproducible evalua-tion. Medical image analysis 63, 101694.

Zhang, Y., Dong, Z., Phillips, P., Wang, S., Ji, G., Yang, J., Yuan, T.-F., 2015. Detection of subjects and brain regions related to alzheimer's disease using 3d MRI scans based on eigenbrain and machine learning. Frontiers in computational neuroscience 9, 66.

Zheng, S., Lu, J., Zhao, H., Zhu, X., Luo, Z., Wang, Y., Fu, Y., Feng, J., Xiang, T., Torr, P.H.S., et al., 2021. Rethinking semantic segmentation from a sequence-to-se-quence perspective with transformers. In: Proceedings of the IEEE/CVF Conference on Computer Vision and Pattern Recognition, pp. 6881–6890.